\documentclass[10pt,twocolumn,letterpaper]{article}

\usepackage{cvpr}              

\usepackage{graphicx}
\usepackage{amsmath}
\usepackage{amssymb}
\usepackage{booktabs}
\usepackage{blindtext}

\usepackage[pagebackref,breaklinks,colorlinks]{hyperref}

\usepackage[capitalize]{cleveref}
\crefname{section}{Sec.}{Secs.}
\Crefname{section}{Section}{Sections}
\Crefname{table}{Table}{Tables}
\crefname{table}{Tab.}{Tabs.}

\def\cvprPaperID{5371} 
\def\confName{CVPR}
\def\confYear{2023}

\begin{document}

\title{Rate-Perception Optimized Preprocessing for Video Coding}

\author{Chengqian Ma, Zhiqiang Wu, Chunlei Cai, Pengwei Zhang,\\ Yi Wang, Long Zheng, Chao Chen, Quan Zhou\\
Bilibili Inc.\\
Shanghai, China\\
{\tt\small \{machengqian01, wuzhiqiang01, caichunlei, zhangpengwei,}\\ 
{\tt\small wangyi, zhenglong, chenchao02, zhouquan\}@bilibili.com}
}

\twocolumn[{%
\renewcommand\twocolumn[1][]{#1}%
\maketitle
\begin{center}
    \centering
    \captionsetup{type=figure}
    \includegraphics[width=\textwidth,keepaspectratio]{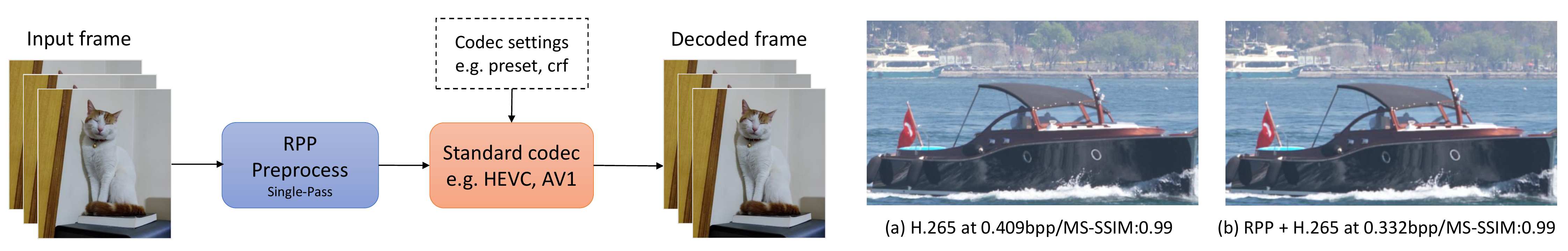}
    \captionof{figure}{\textbf{Left:} The deployment workflow of RPP: single-pass of the input frame and suitable with all standard video codec. \textbf{Right:} Frame segments of H.265 vs. RPP + H.265 at same MS-SSIM. Zoom in to see the details.}
\end{center}%
}]

\begin{abstract}
In the past decades, lots of progress have been done in the video compression field including traditional video codec and learning-based video codec. However, few studies focus on using preprocessing techniques to improve the rate-distortion performance. In this paper, we propose a rate-perception optimized preprocessing (RPP) method. We first introduce an adaptive Discrete Cosine Transform loss function which can save the bitrate and keep essential high frequency components as well. Furthermore, we also combine several state-of-the-art techniques from low-level vision fields into our approach, such as the high-order degradation model, efficient lightweight network design, and Image Quality Assessment model. By jointly using these powerful techniques, our RPP approach can achieve on average, 16.27\% bitrate saving with different video encoders like AVC, HEVC, and VVC under multiple quality metrics. In the deployment stage, our RPP method is very simple and efficient which is not required any changes in the setting of video encoding, streaming, and decoding. Each input frame only needs to make a single pass through RPP before sending into video encoders. In addition, in our subjective visual quality test, 87\% of users think videos with RPP are better or equal to videos by only using the codec to compress, while these videos with RPP save about 12\% bitrate on average. Our RPP framework has been integrated into the production environment of our video transcoding services which serve millions of users every day. Our code and model will be released after the paper is accepted.
\end{abstract}

\section{Introduction}

In recent years, the demand for online streaming high-definition video is growing rapidly, and is expected to continue to grow in the next following years. These streaming high-definition videos cost huge bandwidth. They spend more than 80\% of all consumer Internet traffic \cite{cisco2020cisco}. Therefore, it is essential to build a highly efficient video compression system to generate better video quality at a given bandwidth budget. Thus, many video coding standards have been developed during the past decades, such as H.264 \cite{wiegand2003overview}, H.265 \cite{sullivan2012overview}, H.266 \cite{bross2021overview}, and AOMedia Video 1(AV1) \cite{chen2020overview}. These traditional methods are built on many handcrafted modules, such as block partition, Discrete Cosine Transform (DCT) \cite{1672377}, and intra/inter prediction, etc. While these handcrafted methods have achieved good rate-distortion performance, learned video compression methods \cite{chen2017deepcoder, lu2019dvc, liu2016cu} still attract more and more attention which is inspired by the success of deep neural networks in other fields of image processing. These learned methods claim to achieve comparable or even better performance than traditional codecs. However, most existing learned video compression methods increase the complexity on both the encoder and decoder sides. This computationally heavy decoder makes deployment not viable, especially on end-user devices such as mobile phones and laptops. Some studies try to convert the essential components of standard hybrid video encoder designs into a trainable framework in order to end-to-end optimize all the modules in the video encoder \cite{lu2019dvc, zhang2021dvc}. However, few studies have attempted to use preprocessing methods to improve the rate-distortion performance of video compression systems. 

In this paper, we propose a rate-perception optimized preprocessor (RPP) that can efficiently optimize the rate and visual quality at the same time in an independent single forward pass. In particular, we introduce the adaptive Discrete Cosine Transform (DCT) loss into the training stage of the RPP. In addition, we also engage the full-reference image quality assessment model: MS-SSIM \cite{wang2003multiscale} into the training part to optimize the perceptual quality of the model. At the same time, a light-weight fully convolutional neural network with attention mechanism is designed by us to improve efficiency.

The contributions of our work can be summarized as follows:

\begin{itemize}
    \item We first introduce the adaptive Discrete Cosine Transform (DCT) loss which can reduce spatial redundancy meanwhile still keeping the important high frequency component for the content. From our experiments, involving the adaptive DCT loss in training can significantly save the bit rate and maintain the visual quality of the video.
    
    \item We propose a rate-perception optimized preprocessor (RPP) which is a light-weight fully convolutional neural network with attention mechanism. The RPP model is balanced between perception and distortion by utilizing both adaptive DCT loss and reference-based IQA loss functions. We also introduce the higher-order degradation model into our training stage to enhance the visual quality of the preprocessed frame.
    
    \item Our approach can be easily plugged into the preprocess pipeline of any standard video codec, such as AVC, HEVC, AV1 or VVC. Powered by our approach, these standard video codec can achieve better performance in BD-rate without any changes and sacrifices in video encoding and decoding. Compared with state-of-the-art video codec method, our model can reduce the BD-rate by about 16.27\% in average under multiple quality metrics. Furthermore, our RPP model are extreme efficient which can achieve 1080p@87FPS during the inference which is far beyond real-time efficiency.
\end{itemize}

\section{Related Work}

\subsection{Image Compression}

In the past decades, a lot of traditional image compression methods like JPEG \cite{wallace1991jpeg}, JPEG2000 \cite{christopoulos2000jpeg2000} and BPG \cite{bellard2016bpg} have been proposed. These methods have achieved high performance on reducing the image size efficiently by exploiting hand-crafted techniques. One of the most important parts for those hand-crafted designs is the transformation like DCT. The DCT linearly maps the pixels into the frequency domain. One advantage of the DCT is that it can compact energy which makes it easy to reduce the spatial redundancy of the image. After transformation, these methods quantize the corresponding coefficients and then do the entropy coding. Recently, thanks to the DNN, learning-based image compression methods \cite{balle2016end, balle2018variational, minnen2018joint} have achieved competitive or better performance than the traditional image compression codes.

\subsection{Video Compression}

There is a long history of progress for the video compression methods. During past decades, several video coding standards have been proposed and widely used in the real world, such as H.264 \cite{wiegand2003overview}, H.265 \cite{sullivan2012overview}, H.266 \cite{bross2021overview}, and AOMedia Video 1(AV1) \cite{chen2020overview}. With the continuous development of video coding standards, these traditional video compression methods provided strong performance and made significant improvements. These methods are also practical to use with the hardware support in the real-world applications, such as online video streaming, digital tv, etc. In recent years, a lot of DNN based methods have been proposed for every part of the video coding, such as intra prediction and residual coding \cite{chen2017deepcoder}, mode decision \cite{liu2016cu}, entropy coding, etc. Those methods are employed to improve the performance of one specific module of the traditional video codec. Instead of replacing the particular component of the traditional video compression codec, some approaches focus on the end-to-end optimized video compression framework \cite{lu2019dvc, zhang2021dvc}. In addition, A. Chadha \textit{et al.} \cite{chadha2021deep} tries to converts the essential components of standard video encoder designs into a trainable framework and jointly optimize a preprocessor with the differentiable framework from the end-to-end manner.

\subsection{Metrics}

In the past decades, Peak Signal-to-Noise Ratio (PSNR) was the most widely used full-reference method for assessing video fidelity and quality and it continues to play a fundamental role in evaluating video compression algorithms. However, the PSNR has been proven that has a poor correlation with human perception \cite{girod1993s, wang2009mean}. Thus, a variety of full-reference image quality assessments (IQA) or video quality assessments (VQA) has been proposed \cite{wang2004image, sheikh2006image, li2016toward, barman2018evaluation}. For example, Structural Similarity (SSIM) index \cite{wang2004image} estimates perceptual distortions by considering structural information, and its variant MultiScale-SSIM (MS-SSIM) \cite{wang2003multiscale} provides better performance and more flexibility by incorporating multiscale resolution processing. Video Multi-method Assessment Fusion (VMAF) \cite{li2016toward, li2018vmaf} is another main stream evaluation metric in the real-world industry where lots of famous commercial companies like Netflix \cite{li2018vmaf}, Meta \cite{regunathan2020efficient}, Tiktok \cite{zhang2021video}, Intel \cite{kossentini2020svt} etc., and standardization such as AOMedia \cite{chen2018overview} adopt it for video codec evaluation. VMAF combines three quality features: Visual Information Fidelity (VIF) \cite{sheikh2006image}, Detail Loss Metric (DLM) \cite{5765502}, and Motion, to train a Support Vector Machine (SVM) regressor \cite{cortes1995support} to predict subjective score of video quality. Lot of studies have demonstrated that VMAF is remarkably more correlated to the Mean Opinion Score (MOS) than SSIM and PSNR \cite{barman2018evaluation, rassool2017vmaf, zhang2020comparing}.

\subsection{Image Enhancement}

Image enhancement has been a long-standing problem for its vitally practical value in all kinds of vision applications. Recently, with the development of deep learning techniques such as network design and gradient-based optimization problems, the learning-based methods \cite{Agustsson_2017_CVPR_Workshops, wang2021real} have shown promising performance in various fields of image enhancement including super-resolution, denoising, deblurring, etc. Some methods \cite{hui2019lightweight, liu2020residual} aim at achieving real-time image super-resolution with well-designed lightweight CNN which can obtain better results with limited computational effort. Other approaches \cite{elad1997restoration, wang2021real} focus on designing the degradation models which aim to model the complex degradation process of the image. Wang \textit{et al.} \cite{wang2021real} uses a high-order degradation process to simulate complex real-world degradations. While lots of great works have been done in the image enhancement field, there are rare works that utilize methods and techniques with video coding.

\begin{figure*}
    \centering
    \includegraphics[width=\linewidth,keepaspectratio]{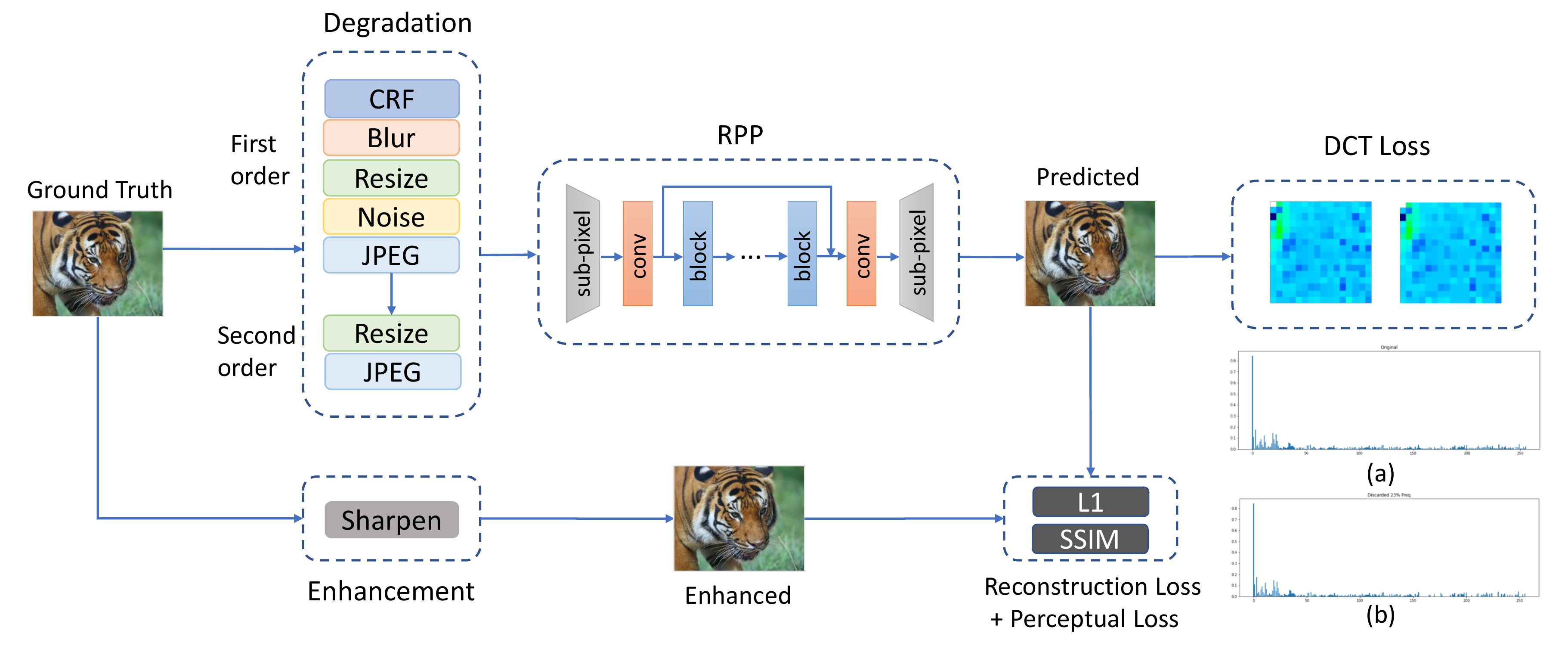}
    \caption{Example framework of training RPP. (a) is the histogram of frequency coefficient of the predicted frame. (b) is the histogram of frequency coefficient filtered by the adaptive DCT function }
\end{figure*}

\section{Proposed Method}

\subsection{Overview}

In this section, we give a brief overview of our rate-perception optimized preprocessing (RPP) method. The goal of our preprocessing model is to provide a preprocessed input frame that is optimized with both rate and perception via a learnable preprocessing neural network. Specifically, in order to optimize our model in the balance between rate and distortion, we design an adaptive DCT loss that can reduce the spatial redundancy and keep the essential high frequency components for perception in the meantime. On the other hand, for the perception optimization part, we aim to perceptually enhance our preprocessed input frame by using the full-reference IQA model: SSIM. We utilize the IQA model as the loss function in our training procedure. In addition, we combine the higher-order degradation modeling process to simulate real-world complex degradation \cite{wang2021real}. By using this higher-order degradation method to generate the pair of training data, our preprocessing network can be trained to handle some complicated degradations in the real world which can also improve the perceptual quality of the output from the network. Furthermore, for the sake of performance and efficiency, we construct a light-weight fully convolutional neural network with a channel-wise attention mechanism \cite{hu2018squeeze}. In the deployment framework, for a given video frame {$\it {f_i}$}, it simply goes a single forward pass through the RPP network. Then the processed frame {$\it {f_o}$} from the RPP network can be encoded by a standard video codec, such as an AVC \cite{wiegand2003overview}, HEVC \cite{sullivan2012overview}, VVC \cite{bross2021overview}, or AV1 \cite{chen2020overview} encoder.

\subsection{Adaptive Discrete Cosine Transform Loss}

Although it has been many years since DCT was first introduced in image/video compression algorithms, because of its high effectiveness and ease of use, DCT-like transforms are still the mainstream transform today. Generally, the basis function of two-dimensional(2D)DCT can be written as: 

\begin{equation}
\begin{aligned}
    & B_{h, w}^{i, j} = cos\frac {h\pi}{H}({i + \frac{1}{2}})cos\frac {w\pi}{H}({j + \frac{1}{2}})
\end{aligned}
\end{equation}
Then the 2D DCT is formulated as:
\begin{equation}
\begin{aligned}
    & F_{h, w} = \sum_{i=0}^{H-1}\sum_{j=0}^{W-1}f_{i, j}B_{h, w}^{i, j}
\end{aligned}
\end{equation}
$$ \it {s.t.\quad h} \in \{0, 1, \cdots, H - 1 \}, {\it w \in \{0, 1, \cdots, W - 1 \}} $$
where $F \in \mathbb{R^{H \times W}}$ is the 2D DCT frequency spectrum, $f \in \mathbb{R^{H \times W}}$is the input frame, $H$ is the height of $f$, and $W$ is the width of $f$. Normally, height and width are the same. $H$ and $W$ are usually denoted as $N$ in most common cases.

With the input of the frame $f$, it converts blocks of pixels into same-sized blocks of frequency coefficients. As we mentioned, the DCT has a crucial property which is that the blocks of frequency coefficients separate the high-frequency components from the low frequency. In an image, most of the energy will be concentrated in the lower frequencies, so in the traditional compression algorithms, they simply throw away the higher frequency coefficients to reduce the spatial redundancy. However, some of the high frequency components also play a very important role in the visual quality of the whole frame. Therefore, we first introduced the adaptive DCT loss for video preprocessing. First, we use DCT to transform a frame $f$ into the frequency domain. Second, we select the frequency coefficients $I$ which belong to the high frequency components by using the ZigZag order traversal. The formula can be written as:
\begin{equation}
F'_{h, w} = F_{h, w} * I_{h, w}
\end{equation}
\begin{equation}
where \quad I_{h, w} = 
    \begin{cases}
        0 & \text{if } \left (h + w \right) < S \\
        1 & \text{if } \left (h + w \right) \geq S
    \end{cases}
\end{equation}
$$ S \in \{0, 1, \cdots, (H - 1)(W - 1) \} $$
In the DCT frequency domain, the value of the frequency coefficient means how much energy is in this frequency component in the whole frame. If a frequency component has less energy, it means that this frequency component is relatively less essential to reconstruct the frame. So we want to throw away some high frequency component with a relatively small value of coefficients. In this case, we do the mean average of the absolute value of these selected coefficients $F'_{h, w}$ to get a Threshold $T$, which can be formulated as:
\begin{equation}
T = \frac{1}{H \cdot W} \sum\limits^{H-1}_{h=i} \sum\limits^{W-1}_{w=j} (\lvert F'_{h, w}\rvert)
\end{equation}
$$ where\quad i + j \geq N$$
 If $|F'_{h, w}|$ is smaller than Threshold $T$, this means it has less effect on reconstructing the frame than the average. Then we select it into another set of the coefficients $F''_{h, w}$. Finally, we calculate the mean absolute error between the filtered DCT frequency coefficients $F''_{h, w}$ and zero, which can be written as:
\begin{equation} \label{eq6}
    \begin{aligned}
    \mathcal L_{dct} & = \sum\limits^{H-1}_{h=i} \sum\limits^{W-1}_{w=j}|F''_{h, w} - 0|, \\
    & F''_{h, w}\in\{\lvert F'_{h, w}\rvert<T\} \quad and \quad i + j \geq N
    \end{aligned}
\end{equation}
 By using this loss function in the model training, the model will be optimized to preserve the essential high frequency components and discard some trivial high frequency components. With this optimization, the frame processed by the model can make the video encoder allocate more bit rates to these important high frequency components such as edges and contrast areas. In the meanwhile, since the adaptive DCT loss function will filter some trivial high frequency components to be zero, it can also benefit the entropy coding process \cite{huffman1952method, 5390830} which will consume much less bitrate with consecutive zeros.

\subsection{Network and Image Degradation}

Inspired by the light-weight network architectures from the image enhancement field, we adopt a few ideas from them \cite{hui2019lightweight, liu2020residual}. Specifically, based on the feature extraction block like RFDB \cite{liu2020residual}, we add the channel attention mechanism \cite{hu2018squeeze} into the block in order to let the network pay more attention to different channel frequencies. Moreover, we use an efficient sub-pixel convolution which is first introduced by Shi \textit{et al.} \cite{shi2016real} to downscale and upscale the resolutions of feature maps. The overall network architecture is shown in Fig. 4.

The way to model the degradation of the training data is important to improve the visual quality during network training. We include some general degradation \cite{elad1997restoration} methods into our degradation model, such as blur, noise, resize, and JPEG compression. For the blur, we model our blur degradation with isotropic and anisotropic Gaussian filters. We choose two commonly-used noise types which is Gaussian noise and Poisson noise for noise degradation. For resizing, we use both upsampling and downsampling with several resize algorithms including area, bilinear, and bicubic operations. Since in the real-world applications, the input frames of our framework mostly are decoded from a compressed video, so we add the video compression degradation which may introduce blocking and ringing artifacts from spatial and time domain. As we mentioned before, High-order degradation modeling \cite{wang2021real} has been proposed to better simulate the complex real-world degradations. We utilize this idea in our image degradation model as well. By generating training pairs with these degradation models, our objective is to make the model have the ability to remove common noise and compression noise, which can also optimize the rate because video codec can not encode the noise well. 

\subsection{Loss Functions}

Our target is to train our preprocessing network by optimizing rate and perception at the same time. In order to perform the optimization of both rate and perception on the reconstructed frame $\hat f$ relative to the input frame $f$, we combine the adaptive DCT loss $\mathcal L_{dct}$, reconstruction loss $\mathcal L_{r}$  and perceptual loss $\mathcal L_{p}$ together to optimize the model. $\mathcal L_{dct}$ is the method introduced by us to optimize the rate and distortion in Eq.\ref{eq6}. For reconstruction loss $\mathcal L_{r}$, we want to ensure the basic reconstruction ability of the model so that we adopt the L1 distance as our reconstruction loss, which can be formulated as:
\begin{equation}
    \begin{aligned}
    \mathcal L_{r} & = \frac{1}{HW}\sum\limits^{H-1}_{i=0} \sum\limits^{W-1}_{j=0}|f^{GT}_{i,j} - \hat f_{i,j}|
    \end{aligned}
\end{equation}
in which $f^{GT}$ is the processed ground truth of the $f$. It is common knowledge that the contrast or edge in the high frequency areas has a higher correlation with human perception. Multiscale structural similarity (MS-SSIM) \cite{wang2003multiscale} is proven by being good at preserving the structural information and contrast in high frequency regions. Thus, we adopt the MS-SSIM as our perceputal loss part, which can be written as:
\begin{equation}
    \mathcal L_{p} = 1 - \mathcal L_{MS-SSIM}(\hat f, f^{GT})
\end{equation}
With the combination of $\mathcal L_{dct}$,$\mathcal L_{r}$ and $\mathcal L_{p}$, our overall loss function can be formulated as:
\begin{equation}
    \begin{aligned}
    \mathcal L_{all} = \lambda_1 \mathcal L_{dct} + \lambda_2 \mathcal L_{p} + \mathcal L_{r}
    \end{aligned}
\end{equation}
Where $\lambda_1$ and $\lambda_2$ are the rate and perceptual coefficients respectively. 

\section{Experiments}

\begin{figure*}[ht]
    \centering
    \includegraphics[width=\textwidth,keepaspectratio]{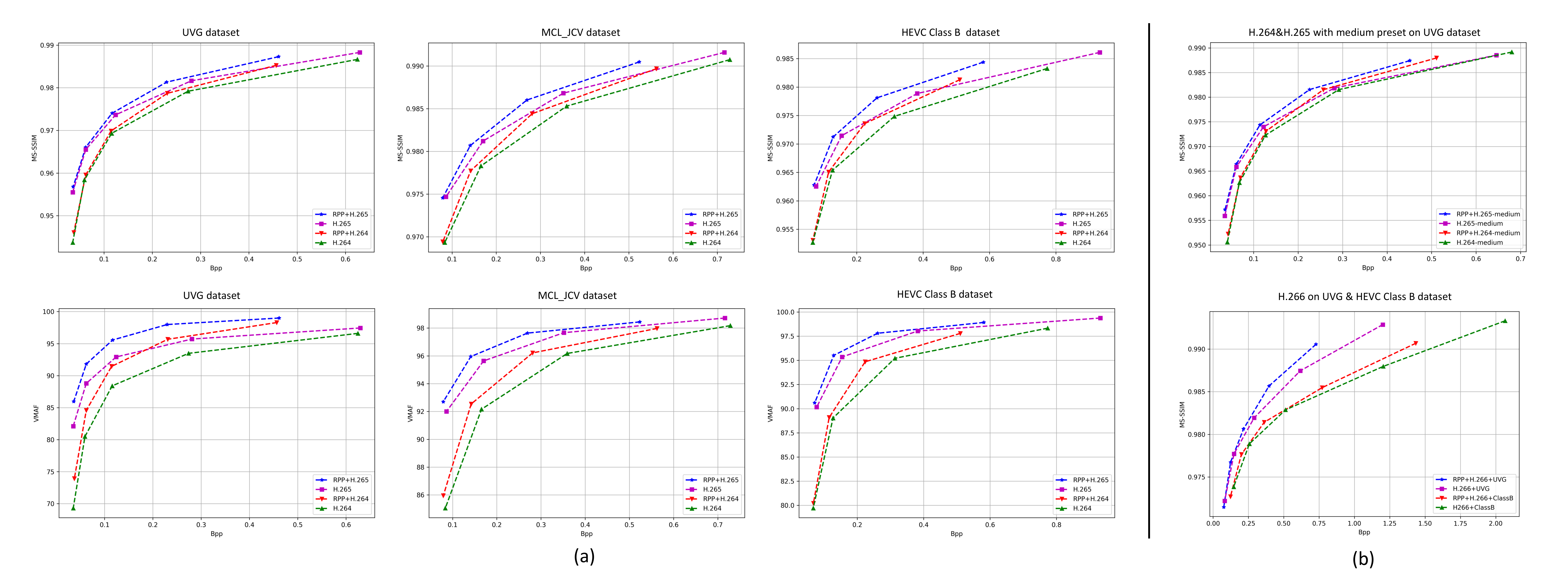}
    \caption{(a) Rate distortion curves for UVG dataset, MCL\_JCV dataset, and HEVC Class B dataset on MS-SSIM and VMAF. Curves are plotted for the standard codec and RPP + standard codec. The corrrsponding BD rates for our proposed method are reported in Tables 1, 2 and 3, repsectively, for each dataset. (b) Top: Rate distortion curves with medium preset for UVG dataset on MS-SSIM. Bottom: Rate distortion curves for H.266 of UVG dataset and HEVC Class B dataset on MS-SSIM.}
\end{figure*}

\subsection{Experiments Setup}

\textbf{Datasets.} We adopt DIV2K and Flickr2K datasets \cite{Agustsson_2017_CVPR_Workshops} for training our RPP model which DIV2K has 1000 high-definition 2K resolution images and Flickr2K has 2650 2K resolution images. To evaluate the performance of our proposed method, we test it on the UVG datasets \cite{mercat2020uvg}, HEVC Standard 1080p Test Sequences \cite{bossen2013common} and MCL-JCV datasets \cite{wang2016mcl}. With the diverse content, these datasets are widely used to evaluate the performance of video compression algorithms.

\textbf{Implementation Details.} We train our RPP model with two stages. The first warm-up stage is that we train the model on reconstruction loss $\mathcal L_{r}$ by using the Adam optimizer \cite{kingma2014adam} with initial learning rate as $1 \times 10^{-3}$, $\beta_1$ as 0.9 and $\beta_2$ as 0.999, respectively. The mini-batch size is set as 32. The resolution of training images is $128 \times 128$ which is randomly cropped from the original images in the datasets. After training 600K iterations with the warm-up training, we use the overall loss function $\mathcal L_{all}$ by setting the $\lambda_1$ as 10, $\lambda_2$ as 0.1 in training, and adjust the learning rate to the $1 \times 10^{-4}$. To be specific in the adaptive DCT loss setting, we use both $N$=8 and $N$=16 to train the network at the same time since the most common size of the macroblock in traditional video codec is 8 and 16. With this setting, we train our RPP model for another 700K iterations so that the model can be converged. The training data of both two training stages are augmented by our two-order image degradation model. The whole training framework is implemented based on Pytorch \cite{paszke2019pytorch} and it takes about only 1 day to train the network by using two NVIDIA GeForce RTX3090. In the deployment stage, the input frame will be first sent into our deployed RPP model to get preprocessed. We set a hyper-parameter here as $\alpha$ to handle the preprocessing intensity of our approach for some cases that do not require intensive preprocessing with our pretrained model setting and are sensitive to all high frequencies information in the video. The value of $\alpha$ is deduced empirically from experiments. The preprocessed frame can be written as: 
\begin{equation} \label{eq10}
    f_p = \alpha f_o + (1 - \alpha) f_i
\end{equation}
where the $f_o$ is the output frame from the RPP model and the $f_i$ is the input frame. Then the preprocessed frame will be encoded by a standard video codec. Importantly, benefitting from our network design, our RPP model can achieve 87.7FPS inference performance for 1080p videos by deployed with TensorRT \cite{vanholder2016efficient} on a single NVIDIA GeForce RTX3090. The inference performance on 720p and 4K is 185FPS and 22.6FPS, respectively.

\textbf{Evaluation Method.} To measure the performance of our proposed method, we use two evaluation metrics: MS-SSIM and VMAF, MS-SSIM is the most common metric in the academic video codec area and VMAF is a mainstream perceptually-oriented metric in the video-streaming industry. We test our proposed method with AVC/H.264, HEVC/H.265 , VVC/H.266, and AV1 which cover all the popular standard video codecs.

\subsection{Experiments Results}

In this section, we show the experimental results of the comparison between standard video codecs and our RPP + standard video codecs. We fix the $\alpha$ = 0.5 in Eq.\ref{eq10} for both HEVC dataset and MCL\_JCV dataset, and $\alpha$ = 1 for UVG dataset. The results of Figure 3(a) and Table 1,2,3 show that our proposed method can obviously improve the BD-rate of both two metrics with standard codecs over all three datasets. The average saving of RPP + H.264 is 18.21\% under VMAF and 8.73\% under MS-SSIM over three datasets.  The average saving of RPP + H.265 is 24.62\% under VMAF and 13.51\% under MS-SSIM over three datasets. Some learning-based video encoders \cite{lu2019dvc} have shown to outperform traditional standard codec only under 'very fast' preset. To demonstrate the generalizability of our approach, We also test our RPP approach with the 'medium' preset. As it shown in the top figure of Figure 3(b), our approach still outperforms the standard codecs which are consistent with the 'very fast' preset results in Figure 3(a). Furthermore, we test our RPP approach with H.266 on UVG dataset and HEVC Class B dataset. As it shown in the bottom figure of Figure 3(b), the average saving of RPP + H.266 is 8.42\% under MS-SSIM over both two datasets. As we expected, our approach can get significant gains when jointly used with all the mainstream standard codecs. In addition, our method has a lower bitrate than the standard codec under the same Quantization Parameter (QP), which can demonstrate the bit-saving ability of our approach.

\begin{table}[ht]
    \captionsetup{justification=centering}
    \centering
    \begin{tabular}{c|cc}
        \hline
          & VMAF & MS-SSIM \\ 
        \hline
        RPP+H.264(veryfast) & -26.92 & -4.86 \\
        RPP+H.265(veryfast) & -39.77 & -8.70 \\
        \hline
        RPP+H.264(medium) & -27.30 & -5.60 \\ 
        RPP+H.265(medium) & -39.24 & -9.58 \\ 
        \hline
    \end{tabular}
    \caption{BD rates on UVG dataset for RPP+H.264 and RPP+H.265 with 'very fast' and 'medium' preset}
\end{table}

\begin{table}[ht]
    \captionsetup{justification=centering}
    \centering
    \begin{tabular}{c|cc}
        \hline
          & VMAF & MS-SSIM \\ 
        \hline
        RPP+H.264 & -15.88 & -9.59 \\
        RPP+H.265 & -19.14 & -11.93 \\
        \hline
    \end{tabular}
    \caption{BD rates on MCL\_JCV dataset for RPP+H.264 and RPP+H.265}
\end{table}

\begin{table}[ht]
    \captionsetup{justification=centering}
    \centering
    \begin{tabular}{c|cc}
        \hline
          & VMAF & MS-SSIM \\ 
        \hline
        RPP+H.264 & -11.84 & -11.75 \\
        RPP+H.265 & -14.94 & -19.90 \\
        \hline
    \end{tabular}
    \caption{BD rates on HEVC ClassB dataset for RPP+H.264 and RPP+H.265}
\end{table}

\begin{figure}[tp]
    \centering
    \includegraphics[width=\columnwidth,keepaspectratio]{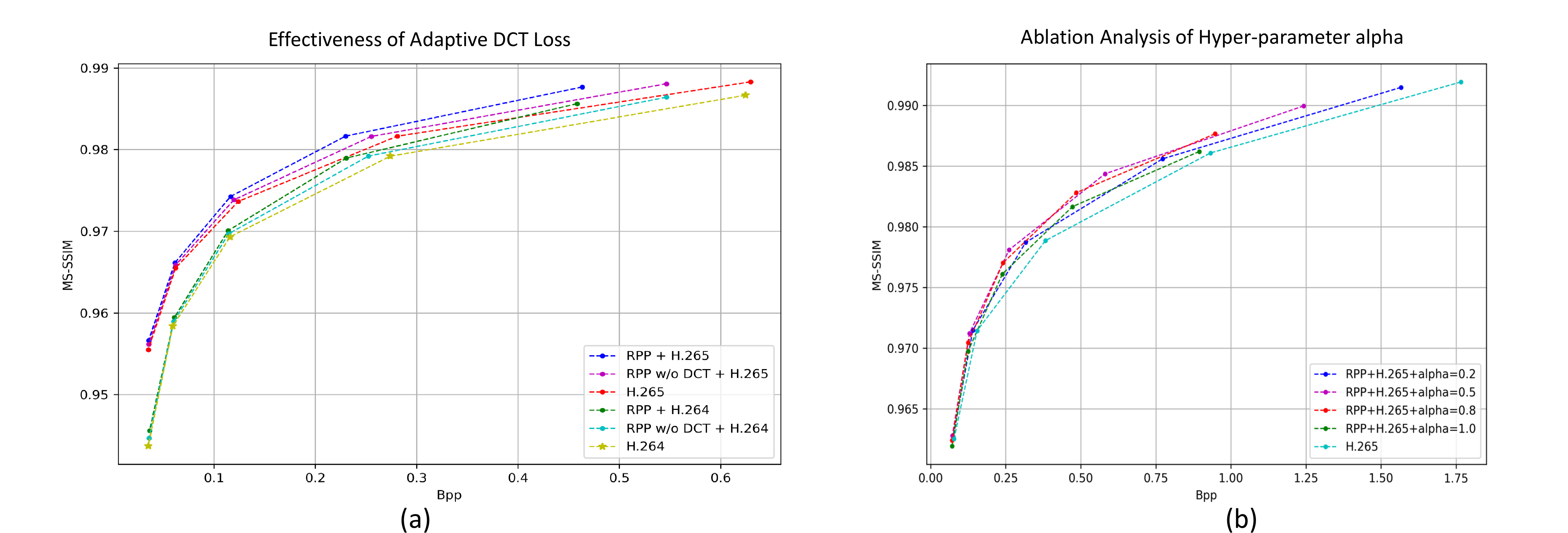}
    \caption{(a) Ablation study of adaptive DCT loss on UVG dataset (b) Ablation study of hyper-parameter $\alpha$ on HEVC Class B and MCL\_JCV dataset}
\end{figure}

\subsection{Ablation Study and Analysis}

\textbf{Effectiveness of Adaptive DCT Loss.} To investigate the effects of the adaptive DCT loss function, we set the $\lambda_1 = 0$ in the $\mathcal L_{all}$ so that the adaptive DCT loss function will not affect the optimization of training. We do this ablation study on the UVG dataset. As shown in Figure 4(a) , we can see that the adaptive DCT loss brings 3.05\% BD-rate saving on H.264 and 6.04\% on H.265 under MS-SSIM, which has a very impressive effect. Compared to the BD-rate savings in Table 1, it contributes over 60\% bitrate savings in the whole approach.

\textbf {Choice and Analysis of Hyper-parameter $\alpha$} We test our approach on HEVC class B dataset and MCL\_JCV by setting different $\alpha$ values (0.2, 0.5, 0.8, 1.0) in Eq.\ref{eq10}. From Figure 4(b), we can see $\alpha$ = 0.5 has the best rate-distortion curve compared to other values of $\alpha$. As we mentioned before, $\alpha$ is to control the preprocessing intensity of our approach. From our perspective, there are two reasons we need to have a hyper-parameter to control the intensity. First, our model is trained at a fixed setting with a small public dataset which means the data is not diverse enough. Second, some videos are extremely sensitive to the high frequency components that our fixed setting pretrained model may over-preprocess.

\section{Conclusion}

In this paper, we propose a rate-perceptual optimized preprocessing (RPP) method to generate a rate-optimized and perceptual-enhanced frame via a neural network for video coding. In the deployment stage, our RPP approach is plug-and-play on the standard video codecs without requiring any changes in encoding and decoding settings. In addition, Our proposed method is also very efficient and can achieve far beyond real-time performance. As shown in experimental results, our RPP approach can achieve considerable and consistent gains with all mainstream standard video codecs on different metrics.

\nocite{*}
{\small
\bibliographystyle{ieee_fullname}
\bibliography{rpp}
}

\end{document}